\documentclass[prd,aps,twocolumn,nofootinbib,superscriptaddress,eqsecnum,floatfix,preprintnumbers,amsmath,amssymb,
nofootinbib,longbibliography,]{revtex4-1}

\usepackage{amssymb,amsmath}
\usepackage{epsfig}
\usepackage[dvipsnames]{xcolor}
\usepackage[utf8]{inputenc}
\usepackage{stmaryrd}
\usepackage{mathrsfs}
\usepackage{mathalfa}
\usepackage{accents}
\usepackage[normalem]{ulem}

\usepackage{graphicx}
\graphicspath{ {c:/Documents/PhysicsNotes/GradResearch/Images/GR_Lens/} }

\newcommand{\pa}{\partial}

\newcommand{\be}{\begin{equation}}
\newcommand{\ee}{\end{equation}}
\newcommand{\bea}{\begin{eqnarray}}
\newcommand{\eea}{\end{eqnarray}}
\newcommand{\ba}{\begin{equation}\begin{aligned}}
\newcommand{\ea}{\end{aligned}\end{equation}}

\newcommand{\beg}{\begin{gather*}}
\newcommand{\eng}{\end{gather*}}
\newcommand{\hh}{,\hspace{0.5cm}}
\newcommand{\hhh}{,\hspace{0.2cm}}

\newcommand{\n}[1]{\label{#1}}

\newcommand{\ts}[1]{{\boldsymbol{#1}}}

\def\XXint#1#2#3{{\setbox0=\hbox{$#1{#2#3}{\int}$ }
\vcenter{\hbox{$#2#3$ }}\kern-.6\wd0}}

\usepackage[makeroom]{cancel}
\usepackage[caption=false]{subfig}
\usepackage[colorlinks=true,
            citecolor=green,
            linkcolor=red,
            filecolor=cyan,
            urlcolor=magenta,
            backref=false]{hyperref}

\newcommand{\ti}[1]{\tilde{#1}}

\newcommand{\by}{\bar{y}}

\begin{document}

\title{Charged Particle Motion Near a Magnetized Black Hole: A Near-Horizon Approximation}

\author{Noah P. Baker}
\email{nbaker@ualberta.ca}
\affiliation{Theoretical Physics Institute, University of Alberta, Edmonton, Alberta, Canada T6G 2E1}
\author{Valeri P. Frolov}
\email{vfrolov@ualberta.ca}
\affiliation{Theoretical Physics Institute, University of Alberta, Edmonton, Alberta, Canada T6G 2E1}

\begin{abstract}
In this paper, the orbits of a charged particle near the event horizon of a  magnetized black hole are investigated.
For a static black hole of mass $M$ immersed in a homogeneous magnetic field $B$, the dimensionless parameter $b=eBGM/ (mc^4)$ controls the radius of the circular orbits and determines the position of the innermost stable circular orbit (ISCO), where $m$ and $e$ are the mass and charge of the particle. For large values of the parameter $b$, the ISCO radius can be very close to the gravitational radius. We demonstrate that the properties of such orbits can be effectively and easily found by using a properly constructed ``near-horizon approximation''. In particular, we show that the effective potential (which determines the position of the orbit) can be written in a form which is invariant under rescaling of the magnetic field, and as a result is universal in this sense. We also demonstrate that in the near-horizon approximation, the particle orbits are stationary worldlines in Minkowski spacetime. We use this property to solve the equation describing slow changes in the distance of the particle orbit from the horizon, which arise as a result of the electromagnetic field radiated by the particle itself. This allows us to evaluate the life-time of the particle before it reaches the ISCO and ultimately falls into the black hole.

\medskip

\hfill {\scriptsize Alberta Thy 4-23}
\end{abstract}

\maketitle

\section{Introduction}

An isolated, non-rotating black hole does not have a magnetic field \cite{GINZBURG}. This is a direct consequence of the famous no-hair theorem \cite{MTW}. However, when a black hole is surrounded by an accretion disk, a magnetic field generated by currents in the disk plasma may exist. This field plays an important role in  black hole astrophysics. For example, the Blandford-Znajek mechanism, which provides an explanation behind the formation of astrophysical jets around spinning supermassive black holes, requires an accretion disk with a strong magnetic field around a rotating black hole
\cite{Blandford:1977ds} (see also \cite{MCKINNEY}  and references therein).

In 2013, evidence of the existence of a magnetic field in the vicinity of the black hole at the center of the Milky way was obtained \cite{Eatough:2013nva,MAGN_MILKY}. This was done by observing the Faraday rotation of the linearly polarized radio waves emitted by the magnetar PSR J1745-2900, which travelled close to the black hole on their way to the Earth. The angle of Faraday rotation of a linearly polarized radio wave beam in magnetized plasma is proportional to the
integral of the product of the longitudinal magnetic field and the density of free electrons over the ray's path. It is also proportional to the wavelength squared \cite{thorne2017modern}. This makes it possible to obtain information about a regular magnetic field by observing the Faraday rotation angle for different wavelengths. The radio waves emitted by the magnetar allow one to measure the magnetic field at a distance from the black hole about $10^5$ of its gravitational radius. According to adopted models, the magnetic field increases inwardly, and the estimated regular magnetic field near the black hole should be about $100G$ \cite{Eatough:2013nva,MAGN_MILKY}.

More recently, in 2021, polarization measurements by the Event Horizon Telescope of the light emitted by the matter surrounding the black hole in the galaxy M87 provided signatures of the magnetic field close to the edge of the black hole \cite{Akiyama_2021}. Based on these observations, the estimated magnetic field strength was shown to be in the range $B\sim 1-30$G.
Estimations based on observations and theoretical models indicate that the magnetic field near stellar mass black holes can in fact be much larger  (see, e.g., \cite{piotrovich2010magnetic,Piotrovich:2014qta}).

Let us emphasize that for realistic magnetic fields expected near black holes, one can neglect their back-reaction on the spacetime geometry. On the other hand, the effect of these fields on the orbits of charged particles near the horizon might be very large. This happens because, in general, the dimensionless quantity $e/(\sqrt{G}m)$ for a particle of charge $e$ and mass $m$ is very large (e.g., for electrons it is of order $10^{16}$). This implies that even close to the horizon of a magnetized black hole, the Lorentz force acting on a charged particle can dramatically change its trajectory. We define a dimensionless parameter which characterizes the strength of the magnetic field, given by
\be
b={eBGM\over mc^4}\, .
\ee
Here, $M$ is the mass of the black hole and $B$ is the characteristic value of the magnetic field in its vicinity.
The meaning of this parameter is as follows.
The surface gravity of a non-rotating black hole of mass $M$ is
\be
\kappa={c^4\over 4GM}\, ,
\ee
which is a red-shifted gravitational field strength calculated at the horizon. Using $\kappa$, one can write the parameter $b$ in the form
\be \n{bbb}
b={1\over 4}{eB\over m\kappa}\, .
\ee
For a relativistic particle of charge $e$ and mass $m$ moving in a magnetic field $B$, the quantity $eB$ is the value of the Lorentz force acting on the particle, while $m\kappa$ characterizes the gravitational force. Relation (\ref{bbb}) shows that the dimensionless parameter $b$ is proportional to the ratio of the Lorentz force and the gravitational one.

There exists a variety of publications discussing different aspects of  magnetized black holes and charged particle motion in their vicinity  (see, e.g., \cite{Prasanna:1978vh,Galtsov:1978ag,Sokolov:1978qb,Sokolov:1978tc,Aliev:1989wx,Aliev:2002nw,Komissarov:2004,BICHAK:2006,Frolov:2010mi,AlZahrani:2010qb,
Strominger:2016,GALTSOV_2018,Kolo__2021,Al_Zahrani_2021,Chakraborty:2022itc,Zahrani:2022fdd,Karas_2023,Hou:2023hto}, along with the texts \cite{Frolov:1998wf,FrolovZelnikov:2011} and references therein).

One interesting observation is that the ISCO of a charged particle near a magnetized black hole can be located much closer to the horizon than ISCO of a neutral particle (see, e.g., \cite{Aliev:2002nw,Frolov:2010mi}). In such a case, black holes (under special conditions) can ``work" as a powerful cosmic particle accelerators (see, e.g., \cite{Frolov:2011ea}).

The present paper is devoted to a detailed investigation of the near-horizon orbits of charged particles in the vicinity of magnetized black holes. Our starting point is a Schwarzschild black hole placed in a homogeneous magnetic field. Such a magnetic field, deformed by the black hole's gravity, can be described by the electromagnetic vector potential, which is directly proportional to the Killing vector generating rotations \cite{WALD}. We focus on particle motion in the equatorial plane. Circular orbits of such a charged particle are fixed points of the reduced Hamiltonian. For a particular direction of motion, the Lorentz force acting on the particle is repulsive and the particle trajectory may lie close to the horizon.

If the proper distance from the horizon $l$ is much smaller than the gravitational radius $r_g$, the curvature effects and the orbit's trajectory bending may be neglected. Put simply, the idea of this paper is that, in such a case, it is sufficient to perform calculations in flat spacetime for the study of near-horizon trajectories. Namely, consider a near-horizon 3-volume with proper length $L\ll r_g$ in both the tangent and orthogonal to the horizon directions. The metric in this domain can be approximated by the Rindler metric. We demonstrate that, by a proper rescaling, one may also define a magnetic field which is static in the Rindler coordinates and which generates the required near-horizon trajectories. In this near-horizon approximation, the Hamiltonian describing the charged particle motion is greatly simplified.

Rindler spacetime is nothing but Minkowski spacetime in specially chosen (Rindler) coordinates.
We demonstrate that in the near-horizon approximation, circular orbits near magnetized black holes map onto so-called ``stationary orbits'' in Minkowski spacetime.

Such stationary curves are well studied and classified (see, e.g., \cite{Letaw}). The geometric properties of these lines are independent of proper time. In particular, the interval between any two points of such a curve depends only on the proper time $s$ between them calculated along the curve. An interesting physical property of stationary worldlines   is the following. If an accelerated Unruh's detector moves along a stationary worldline in the Minkowski vacuum the spectrum of its excitations  is (proper) time independent \cite{Letaw,Korsbakken:2004bv}. The stationary curves are integral lines of the Killing vectors. Another characteristic property of stationary curves is that their Frenet curvatures are constant. As the result  the 4-velocity $\ts{u}$ , 4-acceleration $\ts{w}$ and the "jerk" $\ts{k}=d\ts{w}/ds$ of an observer moving along a stationary worldline have constant components in the Rindler frame. As we shall show this implies that the radiation force acting on a charged particle moving near the horizon is also time independent in the Rindler frame.

It should be emphasized that the magnetic field which is static in the Rindler frame is time dependent in the inertial Minkowski frame. This field is invariant under Rindler boosts and translations in the direction parallel to the horizon. The  charged particle's 4-velocity coincided (up to a normalization constant) with a linear combination of these two vectors.

It is of particular interest that in spite of the fact that we began with the Killing magnetic field in the black hole geometry, the obtained magnetic field close to the horizon and near the equatorial plane is in fact quite general in the following sense:
if the magnetic field respects the imposed symmetry and is regular at the Rindler horizon, it is either constant (in Minkowski time) and homogeneous, or it has the same leading asymptotic as the field obtained in our procedure. This means that one may reasonably expect that our conclusions regarding the properties of close to the horizon orbits may in fact be valid for a more complicated and realistic structure of a regular near-horizon magnetic field.

The paper is organized as follows. In sections~II and III, we re-derive the equations for circular orbits near static magnetized black holes and discuss their properties. In section~IV, we  introduce the near-horizon approximation. In section~V, we apply this approximation in the of study near-horizon trajectories of charged particles. In section~VI, the dynamical evolution of these orbits induced by the emission of electromagnetic radiation is studied. We also evaluate the lifetime of such orbits. Section~VII contains a brief summary along with our discussion. In the appendix, we discuss  the regular at the Rindler horizon magnetic fields respecting the imposed symmetry properties.

In this paper, we use the sign conventions adopted in \cite{MTW} and the Gaussian system of units.

\section{Charged particle motion near magnetized black holes}

\subsection{Charged Particle Motion}

We express the worldline of a particle in the form $x^{\mu}=x^{\mu}(\lambda)$, where $\lambda$ is some parameter specifying the position of the particle.
To describe charged particle motion in curved spacetime and in the presence of an electromagnetic field, we begin with the following action\footnote{Let us note that in our notation, the Lagrangian $L$ and Hamiltonian $H$ have dimensions $[ML/T]$, which differs from the standard dimension of the energy, $[ML^2/T^2]$. This is because the action $S$ is defined as the integral over the proper length $d\lambda$ rather than the proper time $d\lambda/c$. Since in what follows we shall be working with dimensionless quantities,  this difference in definitions of $L$ and $H$ is not important.}
\ba\n{action}
& S=\int d\lambda L\, ,\\
&L={mc\over 2}\left[ \eta^{-1}g_{\mu\nu}{dx^{\mu}\over d\lambda}{dx^{\nu}\over d\lambda}
-\eta \right]+{e\over c}A_{\mu}{dx^{\mu}\over d\lambda}\, .
\ea
We denote by $m$ and $e$ the mass and electric charge of the particle, and by $A_{\mu}$ the 4D electromagnetic field potential.  The parameter $\eta=\eta(\lambda)$ is a Lagrange multiplier which, under a change in parametrization $\lambda\to \tilde{\lambda}(\lambda)$, transforms as
\be
\eta d\lambda=\tilde{\eta}d\tilde{\lambda}\, .
\ee
The above relation ensures that the action $S$ is parametrization invariant. A variation of the action with respect to $\eta$ yields the following constraint equation
\be
g_{\mu\nu}{d{x}^{\mu}\over d\lambda}{d{x}^{\nu}\over d\lambda}+\eta^2=0\, .
\ee
After variations of the action (\ref{action}), one can always set $\eta=1$. For this gauge,
\be\n{prop}
g_{\mu\nu}u^{\mu}u^{\nu}=-1\hh u^{\mu}={dx^{\mu}\over d\lambda}\, .
\ee
The corresponding parameter $\lambda$ has dimensions of length, and it coincides with the proper time multiplied by $c$. From now on, we shall refer to it as the proper time parameter and use this gauge.

Let us define the 4-momentum
\be
p_{\mu}={\pa L\over \pa {x}^{\mu}_{,\lambda}}=mcu_{\mu}+{e\over c}A_{\mu}\, .
\ee
Then the Hamiltonian $H=p_{\mu}u^{\mu}-L$ may be written in the form
\be \n{HH}
H={1\over 2 mc} (\pi_{\mu}\pi^{\mu}+m^2c^2)\, ,
\ee
where $\pi_{\mu}=p_{\mu}-{e\over c}A_{\mu}$. The constraint equation (\ref{prop}) implies
\be\n{pipi}
\pi_{\mu}\pi^{\mu}=-m^2c^2\, ,
\ee
and the Hamiltonian equations of motion are
\be\n{HAM}
{d{x}^{\mu}\over d\lambda}={\pa H\over \pa p_{\mu}}\hh
{d{p}_{\mu}\over d\lambda}=-{\pa H\over \pa x^{\mu}}\, .
\ee
One may check that these equations correctly reproduce the standard equations for particle motion in curved spacetime and in the presence of an electromagnetic field
\be\n{EQUU}
mc {Du^{\mu}\over D\lambda}={e\over c}F^{\mu\nu}u_{\nu}\, ,
\ee
where  $F_{\mu\nu}=2A_{[\nu ,\mu]}$\, . We use the notation $D/D\lambda$ to denote a covariant derivative along the particle worldline.

Let us note that the equations (\ref{HAM}) are invariant under the following scale trasformations
\ba \n{CCCC}
&{x}^{\mu}\to C_1 {x}^{\mu}\hhh \lambda\to C_1 \lambda\, ,\\
&p_{\mu}\to C_2 p_{\mu}\hhh H\to C_2 H\, ,
\ea
where $C_1$ and $C_2$ are constants.

\subsection{Dimensionless Form of the Equations of Motion}

It is more convenient to deal with equations written in dimensionless form. The interval $ds^2$ has dimension of $[length]^2$. We choose  $dx^{\mu}$ to have dimensions of length, and the metric $g_{\mu\nu}$ to be dimensionless. We will introduce a constant scale factor with dimensions of length and use it to obtain dimensionless quantities. In what follows, we shall discuss particle motion in the gravitational field of a static, spherically symmetric black hole. In this case, a natural scale is its gravitational radius
\be
r_g={2GM\over c^2}\, .
\ee
To distinguish the quantities before and after the rescaling, we use a tilde. We thus have
\ba
&d\ti{x}^{{\mu}}=r_g dx^{\mu}\hhh \ti{g}_{{\mu}{\nu}}=g_{\mu\nu}\, ,\\
&d\ti{s}^2=\ti{g}_{{\mu}{\nu}}d\ti{x}^{{\mu}}d\ti{x}^{{\nu}}=r_g^2 ds^2\, ,\\
&ds^2=g_{\mu\nu}dx^{\mu} dx^{\nu}\,  .
\ea

We write the worldline of a particle in the form $\ti{x}^{{\mu}}=\ti{x}^{{\mu}}(\ti{\lambda})$, where $\ti{\lambda}$ is the ``physical" (dimensional) proper-time parameter. Its dimensionless form is given by $\lambda=\ti{\lambda}/r_g$.
For the particle's velocity $\ts{u}$, acceleration $\ts{w}$ and ``jerk" $\ts{k}$, one has
\ba \n{UUEEKK}
&\ti{u}^{{\mu}}={d\ti{x}^{{\mu}}\over d\ti{\lambda}}={dx^{{\mu}}\over d{\lambda}}=u^{\mu}\, ,\\
&\ti{w}^{{\mu}}={D\ti{u}^{{\mu}}\over D\ti{\lambda}}={1\over r_g}{Du^{{\mu}}\over D{\lambda}}={1\over r_g}w^{\mu}\, ,\\
&\ti{k}^{{\mu}}={D\ti{w}^{{\mu}}\over D\ti{\lambda}}={1\over r_g^2}{Dw^{{\mu}}\over D{\lambda}}={1\over r_g^2}k^{\mu}\, ,
\ea
and the equation of motion (\ref{EQUU}) may be written in the form
\be \n{TTWW}
w^{\mu}=f^{\mu}\hh f^{\mu}={e r_g\over mc^2}\ti{F}^{{\mu}{\nu}}u_{\mu}\, .
\ee

\subsection{Motion Near Magnetized Black Holes}

We write the dimensionless Schwarzschild metric in the form
\ba\n{S1}
&d{s}^2=-f d{t}^2+{d{r}^2\over f}+{r}^2d\omega^2\, ,\\
&d\omega^2=d{\theta}^2+\sin^2{\theta} d{\phi}^2\, ,\\
& f=1-1/{r}\, ,
\ea
where $r=\tilde{r}/r_g$ is the dimensionless radius.

This metric has two commuting Killing vectors
\be \n{S2}
\ts{{\xi}}={\partial\over \partial {t}}\hh \ts{{\zeta}}={\partial\over \partial \phi}
\ee
which generate time translations and rotation about the symmetry axis, respectively. The Killing equations
\be \n{S3}
{\xi}_{(\mu ;\nu)}={\zeta}_{(\mu ;\nu)}=0
\ee
imply that
\be \n{S4}
{\xi}^{\mu}_{\ ;\mu}={\zeta}^{\mu}_{\ ;\mu}=0\, .
\ee
Since the spacetime is Ricci flat, its Killing vectors obey the equations
\be \n{S5}
{\xi}^{\mu\nu}_{\ \ ;\nu}={\zeta}^{\mu\nu}_{\ \  ;\nu}=0\, .
\ee
Equations (\ref{S4})--(\ref{S5}) reveal that the  Killing vectors $\ts{\xi}$ and $\ts{\zeta}$ may be viewed as vector potentials of an electromagnetic field $\ts{{A}}$ in the Lorentz gauge satisfying the source-free Maxwell equations. For $\ts{{\xi}}$, one has a weakly charged static black hole, while $\ts{{\zeta}}$ is the potential for a weakly magnetized black hole.

In what follows, we focus on the latter case and write the corresponding vector potential $\ts{{A}}$ in the form
\be\n{AAZET}
{A}^{\mu}\pa_{\mu}={1\over 2}B \pa_{\phi}\, .
\ee
One may check that the corresponding magnetic field at a distance far from the black hole is homogeneous and directed along the axis of rotation $\theta=0,\pi$. For this field, the equation of motion (\ref{EQUU}) may be written in the following dimensionless form
\be
w^{\mu}=f^{\rho}
\ee
where
\be\n{LORF}
f_{\mu}=2b\zeta_{[\nu ,\mu]}u^{\nu}\hh b={eBGM\over mc^4}\, .
\ee
We denote
\be \n{aaaa}
a_{\mu}=b\zeta_{\mu}\, ,
\ee
so that one has $f_{\mu}=2a_{[\nu ,\mu]}u^{\nu}$.
Since the Killing vectors commute, one has ${\cal L}_{{\xi}}a^{\mu}={\cal L}_{{\zeta}} a^{\mu}=0$. This demonstrates that the magnetic field associated with the Killing vector respects the spacetime symmetry.

As mentioned earlier, the magnetic field for the potential (\ref{AAZET}) is parallel to the axis of symmetry $\theta=0,\pi$ at far distances, and its value is constant and equal to $B$. We note that in a more realistic astrophysical setup, the structure of the magnetic field interacting with the ionized plasma of the accretion disk can be more complicated\footnote{
In the appendix, it is shown that in the absence of a monopole magnetic charge, a regular at the horizon magnetic field respecting the spacetime symmetry near the horizon and close to the equatorial plane has the same leading form as the potential (\ref{aaaa}). In this sense, using such a ``toy model" is sufficient for the study of the motion of charged particles near the horizon.
}.

To obtain the dimensionless Hamiltonian, we use the scaling transformations (\ref{CCCC}) and set
\be
C_1=r_g\hh C_2=mc\, .
\ee
After these transformations, the dimensionless Hamiltonian for a charged particle moving in the equatorial plane $\theta=\pi/2$ in the magnetic field (\ref{AAZET}) takes the form
\be
{H}={1\over 2}\left[ -{1\over f}p_t^2+f p_r^2+{1\over r^2}(p_{\phi}-br^2)^2+1\right] \, .
\ee
One may check that the equations of motion for this Hamiltonian are identical to (\ref{EQUU}).
We write here only one of equations of motion, which we shall use later
\be \n{NLV}
{d\phi\over d\lambda}={1\over r^2}(p_{\phi}-br^2)\, .
\ee
The constraint equation (\ref{pipi}) gives ${H}=0$.

\section{Circular orbits in the equatorial plane}

\subsection{The Reduced Hamiltonian}

Since the Hamiltonian $H$ does not depend on $t$ or $\phi$, the momenta $p_t$ and $p_{\phi}$ are integrals of motion. Thus, one has
\be
{d{p}_t\over d\lambda}=-\frac{\partial H}{\partial t}=0\hh {d{p}_{\phi}\over d\lambda}=-\frac{\partial H}{\partial \phi}=0 \, .
\ee
In other words, if a particle begins its motion with some parameters $p_t$ and $p_{\phi}$, it will remain on the surface $p_t$=const and $p_{\phi}$=const in the phase space. We denote these integrals of motion by  $p_t={\cal E}$ and $p_{\phi}=\ell$, and introduce a reduced Hamiltonian
\ba\n{RH}
&{\cal H}={1\over 2} f p_r^2+{\cal V}_r\, ,\\
&{\cal V}_r={1\over 2}\left[{1\over r^2}(\ell-br^2)^2+1-{{\cal E}^2\over f}\right]\, .
\ea

The equations of motion become
\ba\n{fixed}
&{dr\over d\lambda}={\pa {\cal H}\over dp_r}=f p_r\, ,\\
&{d p_r\over d\lambda}=-{\pa {\cal H}\over dr}=-{1\over 2}f' p_r^2-{\cal V}_r'\, ,
\ea
where  a prime denotes a derivative with respect to $r$.

In what follows, we focus solely on circular orbits. In the reduced 2D phase space $(r,p_r)$, they are represented by fixed points of the Hamiltonian flux
\be
{\pa {\cal H}\over \pa r}={\pa {\cal H}\over \pa p_r}=0\, .
\ee

The constraint equation ${\cal H}=0$ at the fixed point implies ${\cal V}_r=0$. This equation defines the energy ${\cal E}$.
The second equation in (\ref{fixed}) gives ${\cal V}_r'=0$, an equation which determines the radius $r$ of a circular orbit for given values of the integrals of motion.

\subsection{Stable and Innermost Stable Circular Orbits}

Calculations give the following expressions for the second derivatives of the Hamiltonian with respect to the canonical coordinates $r$ and $p_r$ at the fixed point
\be
{\pa^2 {\cal H}\over \pa p_r^2}=f\hhh {\pa^2 {\cal H}\over \pa r\pa p_r}=0\hhh
{\pa^2 {\cal H}\over \pa r^2}= {\cal V}_r''\, .
\ee
The condition of stability for a circular orbit reads ${\cal V}_r''>0$. For ${\cal V}_r''<0$, the orbits are unstable. The ``critical radius" where ${\cal V}_r''=0$ determines the position of the innermost stable circular orbit (ISCO).

One may express the equation ${\cal V}_r=0$ in the form
\be\n{EEUU}
{\cal E}^2={\cal U}\hh {\cal U}=f\left[1+{1\over r^2}(\ell -br^2)^2\right]\, ,
\ee
and the two equations ${\cal V}_r'=0$ and ${\cal V}_r''=0$ then imply
\be \n{UUD}
{\cal U}'=0\hh {\cal U}''=0\, .
\ee
The equation ${\cal U}'=0$ defines the radius of the particle's orbit for given values of $\ell$ and $b$, and the equation ${\cal U}''=0$ determines the radius of the ISCO. These equations written in explicit form are
\bea \n{Ur}
& 2b^2r^5-b^2r^4-2\ell b r^2 +r^2-2\ell^2 r+3\ell^2=0\, ,\\
& b^2r^5-r^2+2\ell b r^2+3\ell^2 r-6\ell^2=0\, .
\eea

Using  these equations, it is possible to show  that for a large parameter $b$, the ISCO for particles moving in the direction with positive $\ell$ (corresponding a repulsive Lorentz force) are close to the horizon. In the leading order of $1/b$, one has \cite{Frolov:2010mi}
\bea \n{ISCO}
&r_{ISCO}=1+{1\over \sqrt{3}b}\, ,\n{e1}\\
&\ell_{ISCO}=b+\sqrt{3}\, .\n{e2}
\eea

\subsection{Canonical Transformation}

The kinetic part of  the Hamiltonian (\ref{RH}) depends on the coordinate $r$. It has the form
\be
{1\over 2}{p_r^2\over \mu(r)}\, ,
\ee
where $\mu(r)=f^{-1}$ plays the role of mass and tends to infinity at the horizon. It is possible to make a canonical transformation which transforms the Hamiltonian to a more familiar form. Such new canonical variables are
\be
p=\sqrt{f}p_r\hh \rho=\int_1^r {dr\over \sqrt{f}}\, .
\ee
In these variables, the Hamiltonian takes the form
\be \n{CAN}
H={1\over 2}p^2+{\cal V}_{\rho}\hh {\cal V}(\rho)={\cal V}_r(r(\rho))\, .
\ee
We note that the new canonical coordinate $\rho$ is nothing but the dimensionless proper distance to the horizon.

\section{Near-horizon approximation}

\subsection{Near-Horizon Geometry}

Let us consider the motion of a charged particle close to the horizon of a magnetized black hole with a large parameter $b$. The physical proper distance $\ti{l}$ of the ISCO orbit (\ref{ISCO}) to the horizon is $\sim r_g/\sqrt{b}$. For sufficiently large $b$, its dimensionless version $l=\ti{l}/r_g$, which characterizes the bending of the orbit, is small.
Near the horizon, the curvature ${\cal R}$ is of the order of $r_g^{-2}$. Also in the regime of large $b$, the quantity ${\cal R} l^2\sim 1/b$ characterizing the role of the tidal forces is also small. This means that for our problem, one may neglect the curvature and use a flat-spacetime approximation. Let us discuss this in more detail.

We rewrite the metric (\ref{S2}) in the form
\be
ds^2=-f dt^2+{dr^2\over f}+{r^2\over \cosh^2 z}(dy^2+dz^2)\, ,
\ee
where we have made the following change of coordinates
\be
\phi=y\hh \sin\theta={1\over \cosh z}\, .
\ee
The second relation implies that
\be
{dz\over d\theta}=-\cosh z\, .
\ee
A point $p_0$ where $y=z=0$ corresponds to a point $\phi=0$ in the equatorial plane. At this point, $\ts{e}_y=\pa_y$ is a unit vector in the direction of $\phi$, while $\ts{e}_z=\pa_z$ is a unit vector in the direction of decreasing polar angle. The third unit 3-vector orthogonal to the vectors $\ts{e}_y$ and $\ts{e}_z$ is $\ts{e}_r=f^{1/2}\pa_r$. Together, the vectors $ \{\ts{e}_y,\ts{e}_z,\ts{e}_r\} $ form
a right-handed orthonormal tetrad at $p_0$.

In the vicinity of  $p_0$ and close to the horizon, one has
\be
r=1+q\hhh 0<q\ll 1\hhh |y|\ll 1\hhh |z|\ll 1\, .
\ee
In this approximation, one has
\be \n{mq}
ds^2=-q dt^2+{dq^2\over q}+dy^2+dz^2\, .
\ee
One may check that this is indeed a metric of flat spacetime, where the Rindler horizon is defined by the equation $q=0$.
Let us now  demonstrate how equations (\ref{e1})--(\ref{e2}) may be obtained in using this near-horizon approximation.

\subsection{Near-Horizon Orbits in the Equatorial Plane}

We again consider the motion of a particle in the equatorial plane (where $z=0$) and use a reduced 3D metric
\be
ds^2=-q dt^2+{dq^2\over q}+dy^2\, .
\ee
We also use the following near-horizon approximation for the vector potential of the magnetic field (\ref{aaaa})
\be\n{am}
a_{\mu}=b(1+2q)\delta_{\mu}^{y}\, .
\ee
Let us note that we fix the values of the parameter $b$, which describes the strength of the magnetic field.

It is convenient to introduce new variables
\be
\hat{q}=bq\hhh \hat{p}=p_r/b\hhh \hat{\cal E}=\sqrt{b}{\cal E}\hhh \hat{\ell}=\ell-b
\, ,
\ee
in which the Hamiltonian (\ref{RH}) takes the form
\ba
&{\cal H}={b\over 2}\hat{q}\hat{p}^2+{1\over 2\hat{q}}\left[\hat{\cal U}-\hat{\cal E}^2\right] \, ,\\
&\hat{\cal U}=\hat{q}[(\hat{\ell}-2\hat{q})^2+1].
\ea
At the fixed point of the Hamiltonian, where $\hat{p}=0$,  the constraint equation takes the form
\be
\hat{\cal E}^2=\hat{\cal U}\, .
\ee
For a given $\hat{\ell}$, the coordinate $\hat{q}$ is determined by the equation
\be \n{CIR}
{d\hat{\cal U}\over d\hat{q}}=0\, ,
\ee
while for the innermost stable orbit, one has
\be \n{ISOC}
{d^2\hat{\cal U}\over d\hat{q}^2}=0\, .
\ee
Explicit forms of equations (\ref{CIR}) and (\ref{ISOC}) are given by
 \ba\n{ppzz}
&12\hat{q}^2-8\hat{q}\hat{\ell}+\hat{\ell}^2 +1=0\, ,\\
&\hat{\ell}=3\hat{q}\, .
\ea
Let us emphasize that both the potential $\hat{\cal U}$ and the equations (\ref{ppzz}) written in the variables $(\hat{q},\hat{\ell})$ are invariant under the rescaling of the magnetic field $b$, and in this sense they are universal.

The first of these equations determines a  coordinate distance $q=\hat{q}/b$ of the trajectory of the charged particle from the horizon. One may also solve this equation and find the corresponding integral of motion $\hat{\ell}$ for the trajectory with a given value of $\hat{q}$. One has
\be\n{ppphhh}
\hat{\ell}=4\hat{q}\pm \sqrt{4\hat{q}^2-1}\, .
\ee
\begin{figure}[!htb]%
    \centering
    \includegraphics[width=0.5\textwidth]{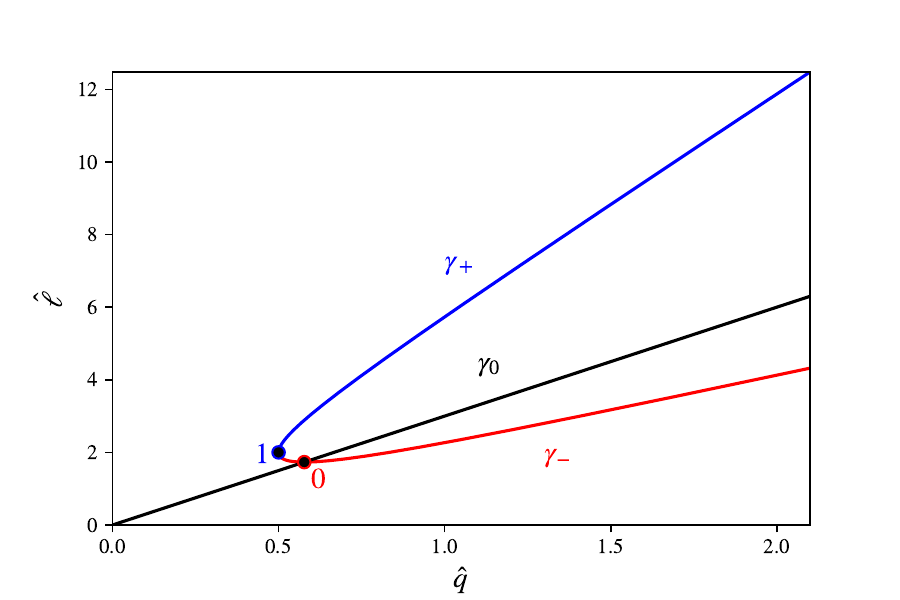}\\[0pt]
    \caption{The parameter $\hat{\ell}$ as a function of $\hat{q}$ as defined by equation (\ref{ppphhh}). The lines $\gamma_{\pm}$ represent the two solutions with the corresponding $\pm$ signs. These lines intersect at point ``$1$''. The
    straight line $\gamma_0$ is the separatice $\hat{\ell}=3\hat{q}$. Above this line, $d^2\hat{\cal U}/d\hat{q}^2<0$, while below it one has $d^2\hat{\cal U}/d\hat{q}^2>0$. Point ``$0$'' represents the innermost stable orbit.
      The part of $\gamma_-$ to the right of point ``$0$'' represents stable orbits.
}
    \label{F1}
\end{figure}
The two branches of this solution for both $\pm$ signs are plotted in Fig.~\ref{F1}, given by the curves $\gamma_{\pm}$. These lines meet at point ``$1$'', where $(\hat{q}=1/2,\hat{\ell}=2)$. The line $\gamma_0$ is the separatice determined by the equation $\hat{\ell}=3\hat{q}$. It divides the $(\hat{q},\hat{\ell})$ plane into two parts. In the upper part of the plane (above $\gamma_0$), $d^2 \hat{\cal U}/d\hat{q}^2<0$, while in the lower part of the plane (below $\gamma_0$),  $d^2 \hat{\cal U}/d\hat{q}^2>0$. The critical point ``$0$'' represents the innermost stable orbit. The separatice $\gamma_0$ intersects $\gamma_-$ at point ``$0$'', where $(\hat{q}=1/\sqrt{3}, \hat{\ell}=\sqrt{3})$. The portion of $\gamma_-$ to the right of this point represents stable orbits.
At this point, the dimensionless radius $r$ and the integral of motion $\ell$ are given by
\be \n{rell}
r=1+{1\over \sqrt{3}b}\hh \ell=b+\sqrt{3}\, .
\ee
These relations correctly reproduce (\ref{ISCO})  in the regime of large $b$.

\section{Particle trajectories}

\subsection{Magnetic Field Near the Rindler Horizon}

\begin{figure}[!htb]%
    \centering
    \includegraphics[width=0.3\textwidth]{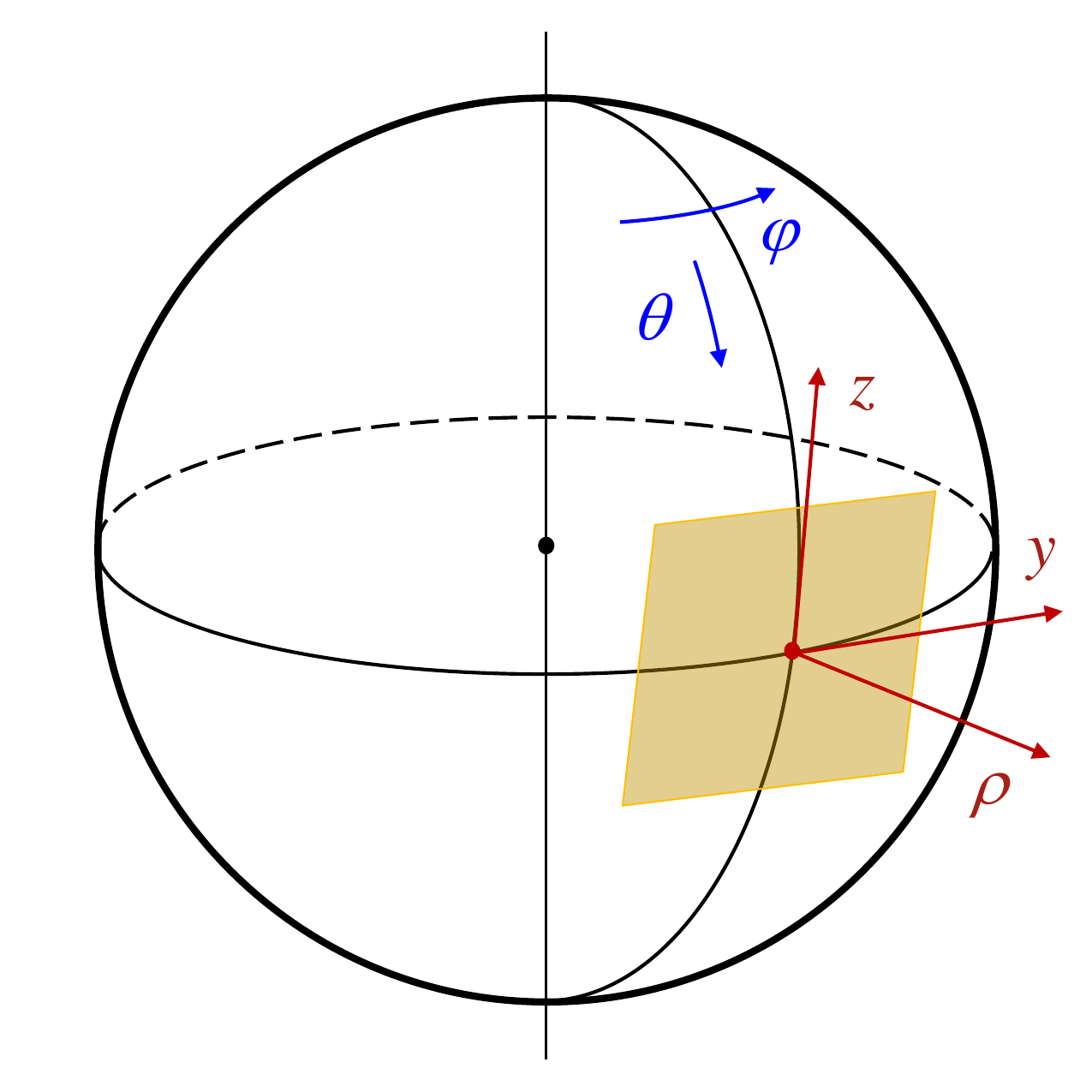}\\[0pt]
    \caption{The near-horizon approximation. The 2D $(y,z)$-plane is tangent to a unit sphere representing the event horizon of the black hole. The $\rho$-axis is orthogonal to the $(y,z)$-coordinate plane. In the adopted near-horizon approximation, the coordinate $\rho$ may be understood as the proper distance from the horizon.}
    \label{F2}
\end{figure}

Let us make the following change of variables
\be
t=2\tau \hh q={1\over 4}\rho^2\, .
\ee
In these coordinates, the metric (\ref{mq}) takes the form
\be\n{Rindler}
ds^2=-\rho^2 d\tau^2+d\rho^2+dy^2+dz^2\, ,
\ee
which is nothing but the standard Rindler form of a flat metric. A schematic outlining the variables used in the near-horizon approximation is shown in Fig.~\ref{F2}. If $(T,X,Y,Z)$ are Minkowski coordinates, then one has
\be
T=\rho\sinh(\tau)\hhh X=\rho\cosh(\tau)\hhh Y=y\hhh Z=z\, .
\ee
The 4-velocity of an observer at rest at $\rho$ in the Rindler frame is
\be \n{OBUU}
u_0^{\mu}={1\over \rho}\delta_{\rho}^{\mu}\, .
\ee
The parameter $\tau$ is simply the dimensionless proper time as measured by an observer at rest at $\rho=1$.

The dimensionless vector potential associated with the Killing vector $\ts{\zeta}$ written in $(\tau,r,y,z)$ coordinates is
\be
a_{\mu}=a_y \delta_{\mu}^y\hh
a_y=b{r^2\over (\cosh z)^2}\, .
\ee
Near the equatorial plane and close to horizon, where $\rho,y\ll1$,  the leading term of the expansion of $a_y$ is of the form
\be \n{RRAA}
a_y=b+{b\over 2}(\rho^2-2z^2)\, .
\ee
The non-vanishing components of the field strength $F_{\mu\nu}=2a_{[\nu ,\mu]}$ for the potential (\ref{RRAA}) are
\be \n{RRFF}
F_{\rho y}=b\rho\hh F_{zy}=-2bz\, .
\ee
Let us note that these expressions were obtained in using the Killing ansatz (\ref{AAZET}) for the magnetic field. In the appendix we demonstrate that, in the absence of a magnetic monopole contribution, expressions (\ref{RRAA}) and (\ref{RRFF}) are nothing but the leading asymptotics for any regular static and $y-$independent magnetic field in the near-horizon approximation.

On the plane at $z=0$, one has $F_{zy}=0$.
The magnetic field vector in the frame of an observer at rest in the Rindler frame is
\be \n{BBB}
B_{\alpha}=-{1\over 2}e_{\alpha\beta\mu\nu}u_0^{\beta}F^{\mu\nu}\, ,
\ee
where $u_0^{\mu}$ is the 4-velocity of the observer (\ref{OBUU}) and $e_{\alpha\beta\mu\nu}$ is the antisymmetric tensor, $e_{0123}=\rho$. The only non-vanishing component of $B_{\mu}$ is
\be\n{BBZZ}
B_z=b\rho\, .
\ee

\subsection{Stationary Worldlines}

In Minkowski spacetime, there exists a special class of worldlines which are called stationary \cite{Letaw}. Their characteristic property is that the geodetic interval between any of its two points
depends only on the proper time interval. The stationary curves in Minkowski spacetime were described and classified by Letaw \cite{Letaw}. He demonstrated that the stationary worldlines are  solutions of the Frenet equations when the curvature invariants are constant. In this subsection, we demonstrate that the worldline of a charged particle moving in the magnetic field (\ref{BBZZ}) is a stationary curve.

For this purpose, let us consider a particle moving in $z=0$ plane and write its worldline $X^{\mu}=X^{\mu}(\lambda)$ in Minkowski coordinates
\be \n{XXX}
X^{\mu}=(\rho_0\sinh(\sigma \lambda),\rho_0\cosh(\sigma \lambda),V\lambda,0)\, .
\ee
As earlier $\lambda$, is the dimensionless proper time parameter along the trajectory. The particle's 4-velocity is then
\be\n{uuuu}
u^{\mu}={dX^{\mu}\over d\lambda}=(\rho_0\sigma\cosh(\sigma \lambda),\rho_0\sigma\sinh(\sigma \lambda),V,0)\, .
\ee
The normalization condition $\ts{u}^2=-1$ implies
\be\n{ssss}
\sigma={\sqrt{1+V^2}\over \rho_0}\, .
\ee
The  curve (\ref{XXX}) is uniquely defined by two independent parameters: $\rho_0$ and $V$.
One may check that the curve in the $(X-Y)$-plane representing the trajectory of this particle is
a catenary
\be\n{XXYY}
X=\rho_0 \cosh\bigg( {\sqrt{1+V^2} \over \rho_0 V} Y \bigg).
\ee

One may also easily find the four-acceleration ${w}^{\mu}=du^{\mu}/d\lambda$ and the ``jerk" ${k}^{\mu}=dw^{\mu}/d\lambda$
\ba
&w^{\mu}=\rho_0\sigma^2(\sinh(\sigma \lambda),\cosh(\sigma \lambda),0,0)\, ,\\
&k^{\mu}=\rho_0\sigma^3(\cosh(\sigma \lambda),\sinh(\sigma \lambda),0,0)\, ,
\ea
which obey the relations
\be
w^{\mu}w_{\mu} = {(1+V^2)^2\over \rho_0^2}\hh k^{\mu}k_{\mu} = -{(1+V^2)^3\over \rho_0^4}\, .
\ee
Let us note that both of these invariants are constant. This implies that the Frenet coefficients calculated for these curves are constant as well, as they should be in the stationary case.

The Rindler coordinates $(\tau,\rho,y,z)$ are related to the Minkowski coordinates $(T,X,Y,Z)$ as follows
\be\n{RTRIN}
\rho=\sqrt{X^2-T^2}\hh \tanh\tau={T\over X}\, .
\ee
These relations give
\ba\n{RRCC}
{\pa \tau\over \pa T}&={X\over \rho^2}\hh {\pa \tau\over \pa X}=-{T\over \rho^2}\, ,\\
{\pa \rho\over \pa T}&=-{T\over \rho}\hh {\pa \rho\over \pa X}={X\over \rho}\, .
\ea
Using the relations
\be
u^{\tau}={\pa\tau\over \pa X^{\mu}} u^{\mu}\hh
u^{\rho}={\pa\rho\over \pa X^{\mu}} u^{\mu},
\ee
one may find $u^{\mu}$ in Rindler coordinates.
One has
\be\n{RRuu}
u^{\mu}=(\sigma,0,V,0)\, .
\ee
This relation shows that $d\tau/d\lambda=\sigma$.

Let us summarize. The particle with the worldline (\ref{XXX}) in Minkowski coordinates remains at a constant distance $\rho=\rho_0$ to the Rindler horizon. For an observer at rest at this distance, the proper time $\tau_0$ is related to $\tau$ via $\tau_0=\rho_0 \tau$. A velocity $v=dy/d\tau_0$ as measured by this observer is
\be
v={V\over \sqrt{1+V^2}}\, .
\ee
The inverse relation determines $V$ as a function of the 3-velocity $v$
\be
V={v\over \sqrt{1-v^2}}\, .
\ee

In a similar way, one finds the components of the vectors $\ts{w}$ and $\ts{k}$ in the Rindler frame
\ba \n{wwkk}
&w^{\mu}=(0,\rho_0\sigma^2,0,0)\, ,\\
&k^{\mu}=(\sigma^3,0,0,0)\, .
\ea

The dimensionless Lorentz force (\ref{LORF}) acting on the charged particle  is
\be\n{FEM}
f^{\mu}=bV\rho \delta_{\rho}^{\mu}\, .
\ee
If the velocity $V$ is directed along the $y$ axis (such that $V$ is positive), the Lorentz force is directed along the $\rho$ axis, i.e., away from the horizon located at $\rho=0$. For motion with a fixed value $\rho=\rho_0$, this repulsive force compensates the ``fictitious" attractive gravity force to the horizon in the Rindler space. This condition reads
\be\n{wwff}
w^{\mu}=f^{\mu}\, .
\ee
Substituting the expression for the 4-acceleration $w^{\mu}$ (\ref{wwkk}) into (\ref{wwff}), one finds
\be\n{ssbv}
\sigma^2=bV\, .
\ee
Using the expression (\ref{ssss}), one obtains
\be\n{brVV}
b\rho_0^2={1+V^2\over V}\, .
\ee
For a given value of $b$, this is the relation between the position of the trajectory $\rho_0$ and the velocity $V$ of the particle.
We have thus shown that a stationary worldline given by (\ref{XXX}) obeying (\ref{ssbv}) does indeed coincide with the trajectory of a charged particle. In the classification of Letaw \cite{Letaw}, such stationary curves belong to class ($\textbf{v}$).

Let us now show that (\ref{brVV}) does in fact reproduce a solution of the first equation in (\ref{ppzz}).
Equation (\ref{NLV}) allows one to write
\be
\hat{\ell}=V+2\hat{q}\, ,
\ee
where  $\hat{q}=b\rho_0^2/4$. Substituting this expression into (\ref{ppzz}) and solving the obtained equation, one finds
\be \n{HHVV}
\hat{q}={1+V^2\over 4V}\, .
\ee
This equation is equivalent to (\ref{brVV}). This means that the orbits which were found by solving the dynamical equations of motion correctly reproduce the results obtained by means of equation (\ref{ppzz}).

Relation (\ref{HHVV}) shows that $d\hat{q}/dV>0$ for $V>1$, where $\hat{q}>1/2$. This means that when $\hat{q}$ decreases (and a point representing the state of the system moves to the left along the lower branch of $\gamma_-$), the velocity $V$ decreases as well. This behavior continues until the critical point $\hat{q}=1/\sqrt{3}>1/2$, at which point the system becomes unstable. At the critical point corresponding to the innermost stable orbit, one has
\be   \n{CR}
V=V_{cr}={1\over \sqrt{3}}\hhh
v_{cr}={1\over 2}\hhh \rho_{cr}={2\over  3^{1/4} \sqrt{b}}\, .
\ee

\subsection{Metric in a Co-Moving Frame}

The equation of motion of the particle in the Rindler frame (\ref{Rindler}) frame is given by
\be
\rho=\rho_0\hh y=V\lambda={\rho_0 V\over \sqrt{1+V^2}}\tau\, .
\ee
To obtain the  metric in a frame co-moving with the particle, it is sufficient to make the following coordinate transformation (shifting to a reference frame with velocity $V$ in the $y-$direction)
\be\n{yyVV}
y={\rho_0 V\over \sqrt{1+V^2}}\tau+\by\, .
\ee
In the coordinates $(\tau,\rho,\by,z)$, the metric is
\begin{align}  \n{VM}
ds^2&=-\left( \rho^2-\rho_0^2{V^2\over 1+V^2}\right) d\tau^2+{2\rho_0 V\over \sqrt{1+V^2}} d\tau d\by \nonumber\\ &+d\rho^2+d\by^2+dz^2\, .
\end{align}
We again emphasize that the coordinate $\tau$ in this expression is the proper time for an observer at rest at $\rho=1$ in the Rindler frame.

For a charged particle moving in a magnetic field, the parameters $\rho_0$ and $V$ are not independent. Using the relation (\ref{brVV}), one can express (\ref{yyVV}) as follows
\be
y=\sqrt{V\over b}+\by\, ,
\ee
and the metric (\ref{VM}) in the co-moving frame takes the form
\begin{align} \n{MOVMET}
&ds^2=-(\rho^2-{V\over b})d\tau^2+2\sqrt{V\over b} d\tau d\rho \nonumber\\
&+d\rho^2+d\by^2+dz^2\, .
\end{align}
We note that this metric, associated with the moving particle, is stationary and does not depend on $\tau$. This is a direct consequence of the fact that its worldline is stationary.

\subsection{Orbital Oscillations}

So far, we have focused mainly on stationary orbits with fixed values of $\rho=\rho_0$. Let us briefly discuss small perturbations of these orbits in the radial direction. For this purpose, we use the Hamiltonian (\ref{CAN}). In the near-horizon approximation, it takes the form
\ba
&{\cal H}={1\over 2}p^2+{\cal V}_{\rho}\, ,\\
&{\cal V}_{\rho}={1\over 2}\left[
(\hat{\ell}-{1\over 2}b\rho^2)^2+1-{4 {\cal E}^2\over \rho^2}
\right]\, .
\ea
Denote by $\rho_0$ the proper distance of the unperturbed orbit. Then expanding the potential ${\cal V}_{\rho}$ near this point, one has
\ba
&{\cal H}={1\over 2}p^2+{\omega^2\over 2}(\rho-\rho_0)^2\, ,\\
&\omega^2=b\left[
2\sqrt{b^2\rho_0^4-4}-b\rho_0^2
\right]\, .
\ea
This expression demonstrates that a radial perturbation of the orbit results in small oscillations around $\rho=\rho_0$.
For a stable orbit, the frequency $\omega$ is real, and it vanishes at the critical innermost stable orbit. For  $\rho_0<\rho_{cr}$, $\omega$ is imaginary, as expected for unstable orbits.

\section{Radiation friction}

\subsection{Electromagnetic Radiation of a Charged Particle}

An accelerated charged particle radiates electromagnetic waves, which reduces its total energy and momentum.
As a result, its motion is modified via an effect called ``radiation damping". We assume that this effect is small, and evaluate how it affects the particle's motion near the horizon. For this purpose, we use the approach developed in the previous sections.
An approximation of the radiation force when the effect is small is given by \cite{Landau1980Classical}
\be\n{ggku}
\ti{g}^{\mu} = {2e^2\over 3c}\bigg(\ti{k}^{\mu}  + \ti{u}^{\mu}\ti{u}^{\nu}\ti{k}_{\nu} \bigg)\, .
\ee
In the presence of radiation, the force $\ti{g}^{\mu}$ should be added to the right-hand side of the equation of motion of the charged particle (\ref{EQUU}). After making a transformation to dimensionless quantities, this equation takes the form
\be
w^{\mu}=f^{\mu}+f_{R}^{\mu}\, ,
\ee
Here, $f^{\mu}$ is the dimensionless Lorentz force given by (\ref{FEM}) and $f_{R}^{\mu}$ is a radiation-friction force
\be\n{ffRgg}
f_R^{\mu}={\epsilon \over b}g^{\mu}\hh g^{\mu}=k^{\mu}  + u^{\mu}u^{\nu}k_{\nu}\, ,
\ee
where $\epsilon$ is a new dimensionless parameter
\be
\epsilon={e^3B\over 3m^2 c^4} \, .
\ee
In what follows, we assume that $\epsilon$ is small.

In order to study how the emitted radiation changes the distance $\rho$ of the charged particle from the Rindler horizon, we consider the radiation force $f_R^{\mu}$ as a perturbation. As such, we keep only terms up to first order in $\epsilon$ in the equations of motion. Since this force is already of order $\epsilon$, it is sufficient to
substitute the unperturbed values of ${k}^{\mu}$ and $u^{\mu}$ into the expression for $g^{\mu}$.
Using the expression (\ref{wwkk}) for $k^{\mu}$ and (\ref{RRuu}) for $u^{\mu}$ in $(\tau,\rho,y,z)$ coordinates, one obtains the following expression for ${f}_R^{\mu}$
\be \n{gggg}
{f}_R^{\mu}={\epsilon \over b}(-{V^2(V^2+1)^{3/2}\over \rho^3},0, -{V(V^2+1)^2\over \rho^2},0 )\,.
\ee
One can see that the $\rho$ and $z$ components of this force vanish, while the other components, $g^{\tau}$ and $g^{y}$, are negative. This implies that as a result of the action of this force, the integrals of unperturbed motion$-$the energy and the $y$-momentum$-$slowly decrease.

\subsection{Evolution of the Orbit}

Let us now discuss how the the electromagnetic radiation of the charged particle affects the orbit of a charged particle moving near the horizon. For a given magnetic field $b$ and in the absence of radiation, the distance $\rho$ of the particle
from the horizon is a function of its velocity. If the radiation is weak, the parameter $\rho$ as well as the velocity $V$ slowly change in time. We choose the velocity $V$ as a parameter describing the state of the system, and assume that $\rho$ is determined by it, i.e., that $\rho=\rho(V)$. We also write
\be \n{EQVV}
{dV\over d\lambda}=\epsilon f(V)\, ,
\ee
where $f(V)$ is a function which characterizes the rate of change of the velocity due to the radiation.
One has
\be
{d\rho\over d\lambda}=\epsilon f {d\rho\over dV}\, .
\ee
The 4-velocity  of the particle moving along a worldline with a slowly changing velocity $V$ is
\be
u^{\mu}=(\alpha,\epsilon f {d\rho\over dV},V,0)\, ,
\ee
and the standard normalization condition $\ts{u}^2=-1$ determines the parameter $\alpha$
\be
\alpha= {1\over \rho} \sqrt{1+V^2}\, .
\ee
Here and later we neglect quantities of the second and higher order in $\epsilon$.

Calculations give the 4-acceleration of the particle $w^{\mu}$ in Rindler coordinates
\ba
&w^{\mu}=(w^{\tau},\alpha^2 \rho,\epsilon f,0)\, ,\\
&w^{\tau}={\epsilon f\over \rho^2 \sqrt{1+V^2}}\left[V\rho+(1+V^2){d\rho\over dV}\right]\, .
\ea
One may check that $u_{\mu}w^{\mu}=0$ as required.
One also finds the following expression for the dimensionless Lorentz force $f^{\mu}$
\be
f^{\mu}=(0,b\rho V,-\epsilon b\rho f (d\rho/dV),0)\, ,
\ee
while the friction force $f^{\mu}_R$ is given by (\ref{gggg}).

Let
\be
J^{\mu}=w^{\mu}-f^{\mu}-f^{\mu}_R\, ,
\ee
then the equations of motion read $J^{\mu}=0$.
The equation $J^{\rho}=0$ implies
\be \n{rrVV}
b\rho^2={1+V^2\over V}\, .
\ee
This relation is identical to the relation (\ref{brVV}). This means that, at least in the leading order, the relation between $\rho$ and $V$ remains the same as in the absence of radiation. The equation $J^y=0$ gives
\be\n{ffff}
f=-{V(1+V^2)^2\over b \rho^2 [1+b\rho (d\rho/dV)]}\, .
\ee
Substituting the expression for $\rho$ which follows from (\ref{rrVV}), one obtains
\be\n{ffVV}
f=-{2V^4(1+V^2)\over 3V^2-1}\, .
\ee
The equation $J^z=0$ is trivially satisfied. One can also check that the last equation, $J^{\tau}=0$, is valid as a result of the relations (\ref{rrVV}) and (\ref{ffVV}).

\begin{figure}[!htb]%
    \centering
    \includegraphics[width=0.45\textwidth]{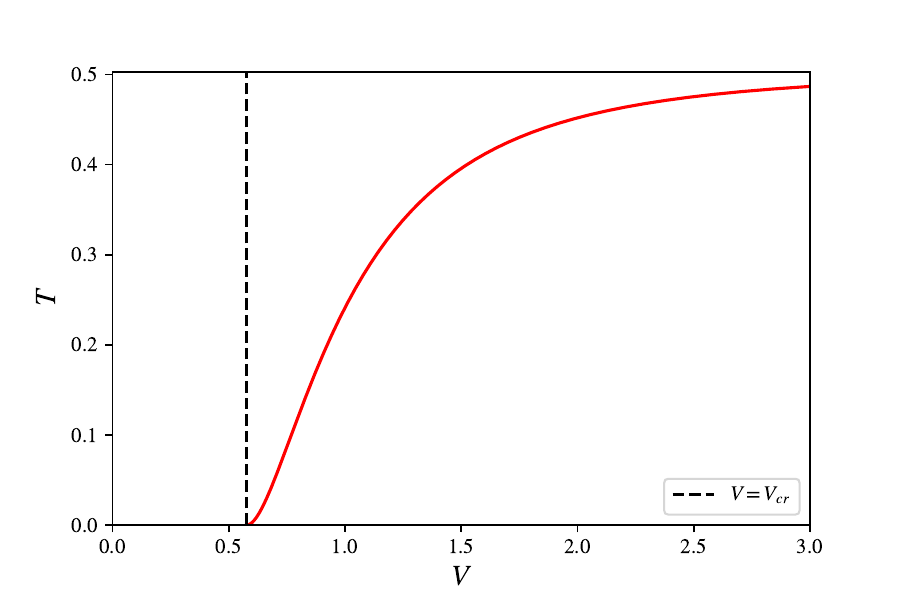}\\[0pt]
    \caption{The function $T=T(V)$. }
    \label{F3}
\end{figure}

A solution of equation (\ref{EQVV}) for $V(\lambda)$ can be written in the following parametric form
\be
\int{dV\over f}=\epsilon \lambda.
\ee
Performing the integration, one gets
\ba
&\int{dV\over f}={\cal T}(V)\, ,\\
& {\cal T}(V)=2\arctan(V)-{1\over 6V^3}+{2\over V}\, .
\ea
We define
\be
{\cal T}_c={\cal T}(V_c=1/\sqrt{3})={\pi\over 3}+{3\sqrt{3}\over 2}\, ,
\ee
and denote
\be \n{TTVV}
T=T(V)={\cal T}_c-{\cal T}(V)\, .
\ee
A plot of the function $T(V)$ is shown in Fig.~\ref{F3}.
The quantity $T/\epsilon$ is the dimensionless proper time parameter $\lambda$ required for a particle with initial velocity $V$ to reach the innermost stable orbit.

\begin{figure}[!htb]%
    \centering
    \includegraphics[width=0.45\textwidth]{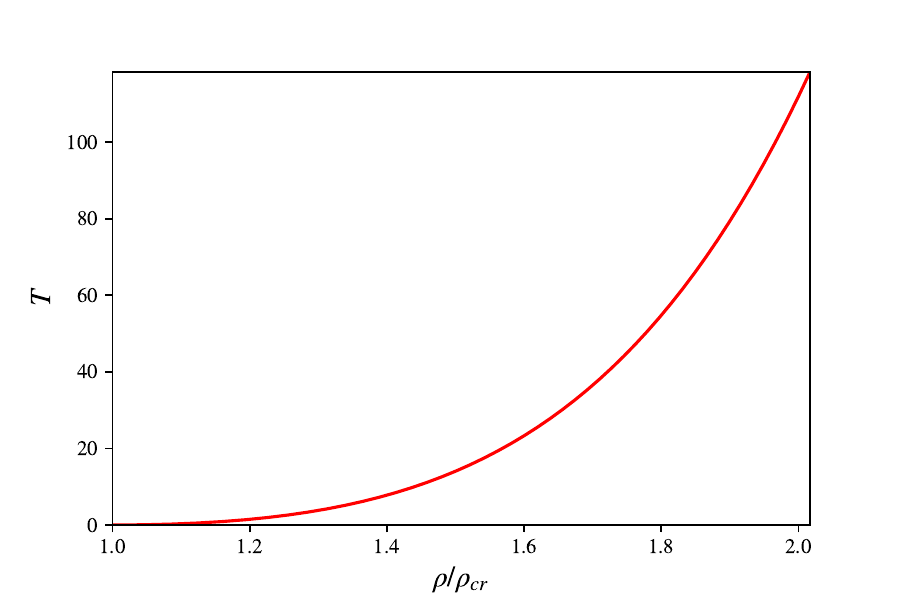}\\[0pt]
    \caption{The lifetime $T$ as a function of $\rho/\rho_{cr}=3^{1/4}\sqrt{b}\rho/2$. }
    \label{F4}
\end{figure}

Suppose a charged particle is initially in an orbit with a given parameter $\rho$. Its velocity is then
\be\n{VVRR}
V=\frac{1}{2}\bigg[b\rho^2-\sqrt{b^2\rho^4-4}\bigg]\, .
\ee
If one substitutes this expression into (\ref{TTVV}), one obtains the lifetime of the charged particle moving near the horizon as a function of its initial position. This function is shown in Fig.~\ref{F4}.

\subsection{Motion at a Fixed Distance to the Horizon}

\begin{figure}[!htb]%
    \centering
    \includegraphics[width=0.45\textwidth]{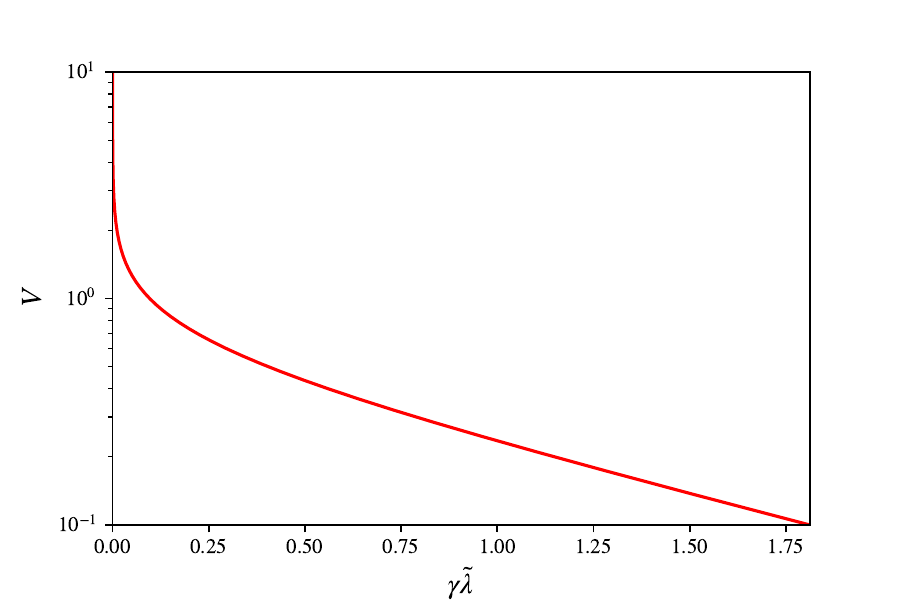}\\[0pt]
    \caption{The velocity $V$ as a function of $\gamma\ti{\lambda}$ for particle motion at a fixed $\rho=\rho_0$. }
    \label{F5}
\end{figure}

As we saw in the previous section, a change in the velocity $V$ leads to a change in the distance parameter $\rho$. Let us now briefly consider a different problem. Namely, we assume that instead of a magnetic field supporting the particle, there exists some other force which is independent of the particle velocity $V$ yet fixes the distance of the particle to the horizon, $\rho=\rho_0$. One then has (cf. \cite{SUEN})
\be
{dV\over d\lambda}=-{2e^2\over 3 m c^2r_g\rho_0^2} V(1+V^2)^2\, .
\ee
Using the physical proper time parameter $\tilde{\lambda}=r_g\lambda$ and dimensional proper distance to the horizon $l=r_g \rho_0$, one may write this relation in the form
\be
{dV\over d\ti{\lambda}}=-\gamma  V(1+V^2)^2\hh \gamma={2e^2\over 3mc^2 l^2}
\, .
\ee
This equation can be easily integrated and its solution may be written in the following parametric form
\be\n{SSS}
\gamma\ti{\lambda}={\cal F}\hh {\cal F}=-\ln V+{1\over 2}\ln(1+V^2)-{1\over 2(1+V^2)}\, .
\ee
Using this relation, one can determine how the velocity $V$ changes with the proper time parameter.
The curve in Fig.~\ref{F5} shows the dependence of the velocity $V$ on the parameter $\gamma\ti{\lambda}$.

The function ${\cal F}(S)$ is monotonically decreasing. Its asymptotics, $V=\infty$ at $\ti{\lambda}=0$ and $V=0$ at $\ti{\lambda}=\infty$, can be easily obtained from (\ref{SSS}).  For small $V_0$, one has
\be
V=V_0 \exp(-\gamma \tilde{\lambda})\, .
\ee
Thus for a small initial velocity $V_0$, the  velocity $V$ decays exponentially with a characteristic time of decay given by
$\gamma^{-1}$.

\section{Discussion}

In this paper, we have discussed the motion of charged particles near magnetized, non-rotating black holes.
In particular, we have studied particles moving in the equatorial plane orthogonal to the magnetic field. We've focused on orbits with a direction of motion corresponding to an outwardly directed Lorentz force  from the black hole horizon.
When the magnetic field parameter $b$ is large, such circular orbits are located close to the gravitational radius $r_g=2GM/c^2$. In this regime curvature and trajectory bending effects can be neglected and one can use the near-horizon approximation.

We considered a near-horizon spatial domain which extends from the horizon up to the proper distance $L\ll r_g$, and assumed that its length in the directions parallel to the horizon also have size $L$.
In this domain, it is shown that one can approximate the exact Schwarzschild geometry by the Rindler metric. This defines a map from the near-horizon geometry to a corresponding Minkowski spacetime. We demonstrated that it is possible to accompany this map with a proper rescaling of the electromagnetic field, such that the near-horizon trajectories of the charged particles map to stationary curves. These timelike curves in Minkowski spacetime have remarkable properties. The interval of a straight line connecting any two points on such a worldline depends only on the proper time interval between these points.
Stationary curves are integral lines for a linear combination of the boost and rotation Killing vectors, and their Frenet curvatures are constant \cite{Letaw}. In the near-horizon approximation, the Hamiltonian of the charged particle is greatly simplified.

As our model, we used a black hole immersed into a homogeneous at infinity magnetic field. However, for the description of the near-horizon motion of charged particles, only its near-horizon asymptotic is important. In fact, as it is shown in the appendix, this near-horizon field is a general asymptotic of \emph{any} regular at the horizon magnetic field respecting the imposed symmetries.

Using the adopted near-horizon approximation, we solved the equations describing the slow evolution of a circular orbit of the charged particle induced by its electromagnetic radiation (see also \cite{GALTSOV_2018}). As a result of the radiation emitted by the particle, it loses both its angular momentum and energy, and the radius of the orbit decreases until the particle reaches the innermost stable orbit and falls into the horizon. We calculated the corresponding lifetime for this process. Such a friction effect on a particle moving near the horizon of a black hole and parallel to its surface has a natural explanation in the framework of the membrane paradigm \cite{MEMBRANE,SUEN}. The motion of the particle creates an electric current on the stretched horizon, and its Ohm's power dissipation is responsible for the friction force acting on the particle. It is quite interesting that a similar effect is known to occur in ``standard" electrodynamics in flat spacetime. Namely, when a charged particle moves parallel to the surface of a conducting material at some given distance, there exists a force which decreases its velocity \cite{BOYER}.

The proper distance of the innermost stable orbit to the horizon depends on the mass of the charged particle. Denote by $\rho_e$ and $\rho_p$ the corresponding distances for the electron and proton then
\be
\rho_p-\rho_e=(\sqrt{m_p/m_e}-1)\rho_e\approx 42 \rho_e \, .
\ee
This implies that for a two component neutral gas of weakly interacting protons and electrons  there may exist a net electric charge associated with an excess of electrons located in the ring of the equatorial plane between $\rho_p$ and $\rho_e$. If one ejects such a neutral gas at some distance from the horizon, the times required for the protons and electrons within it to reach the innermost stable orbit are different. For steady accretion of the neutral plasma into the black hole, this may result in the black hole becoming charged.

For a non-rotating black hole, close to the horizon orbits for large values of the parameter $b$ are ``supported" by the magnetic field. For rapidly rotating black holes, such orbits also exist as a result of the interaction of black hole spin with the particle angular momentum. It would be interesting to extend the near-horizon approximation adopted in this paper to the study the near-horizon circular orbits in Kerr geometry.

\section*{Acknowledgments}

The authors thank the Natural Sciences and Engineering Research Council of Canada and the Killam Trust for their financial support.

\appendix

\section{Electromagnetic Field Near the Rindler Horizon}

In this appendix, we discuss properties of the electromagnetic field near the Rindler horizon. We write the Rindler metric in the form
\be
ds^2=-\rho^2 d\tau^2+d\rho^2+dy^2+dz^2\, .
\ee
The relations between the Rindler coordinates $(\tau,\rho,y,z)$ and the Minkowski coordinates $(T,X,Y,Z)$ are given in (\ref{RTRIN}). Relations (\ref{RRCC}) give the partial derivatives of the Rindler coordinates $\tau$ and $\rho$ with respect to $T$ and $X$.

We define two Killing vectors
\be
\xi^{\mu}\pa_{\mu}=\pa_{\tau}\hh \zeta^{\mu}\pa_{\mu}=\pa_{y}\, ,
\ee
and consider the electromagnetic field $A_{\mu}$ which respects these symmetries
\be \n{LLAA}
{\cal L}_{\xi}A_{\mu}={\cal L}_{\zeta}A_{\mu}=0\, .
\ee
Here, ${\cal L}_{\xi}$ is the Lie derivative along a vector field $\xi$.
Relations (\ref{LLAA}) imply that
\be \n{SSAA}
A_{\mu}=A_{\mu}(\rho,z)\, .
\ee
We impose the Lorenz gauge condition $A^{\mu}_{\ ;\mu}=0$, which for the potential (\ref{SSAA}) in Rindler coordinates takes the form
\be
{1\over \rho}\pa_{\rho}(\rho A_{\rho})+\pa_z A_{z}=0\, .
\ee
Let us denote
\ba
&F_{\mu\nu}=\pa_{\mu}A_{\nu}-\pa_{\nu}A_{\mu}\, ,\\
&J^{\mu}=F^{\mu \nu}_{\ \ ;\nu}\equiv {1\over \sqrt{-g}}\pa_{\nu}\left(\sqrt{-g}F^{\mu\nu}\right)\, .
\ea
Then the source-free Maxwell equations $J^{\rho}=J^z=0$  for the field (\ref{SSAA}) are
\be
\pa_z F_{\rho z}=\pa_{\rho}(\rho F_{\rho z})=0\, .
\ee
A solution of these equations is
\be \n{FRZ}
F_{\rho z}={C\over \rho}\, .
\ee
The other two Maxwell equations, $J^{\tau}=J^{y}=0$, take the following form
\ba
& \rho\pa_{\rho}\left({1\over \rho} \pa_{\rho}A_{\tau}\right)+\pa^2_z A_{\tau}=0\, ,\\
&\pa^2_{\rho}A_y+{1\over \rho}\pa_{\rho}A_y+\pa^2_z A_{y}=0\, .
\ea
Both of these equations allow a separation of variables and their general solutions are
\ba \n{SOL}
&A_{\tau}(\rho,z)=\int_{-\infty}^{\infty} d\omega e^{i\omega z}\tilde{A}_{\tau}(\rho|\omega)\, ,\\
&\tilde{A}_{\tau}(\rho|\omega)=\rho\left[ C^1_{\tau}(\omega) J_1(\omega \rho)+C^2_{\tau}(\omega) K_1(\omega \rho)
\right]\, ,\\
&A_{y}(\rho,z)=\int_{-\infty}^{\infty} d\omega e^{i\omega z}\tilde{A}_{y}(\rho|\omega)\, ,\\
&\tilde{A}_{y}(\rho|\omega)=C^1_{y}(\omega) J_0(\omega \rho)+C^2_{y}(\omega) K_0(\omega \rho)\, .
\ea
Here, $J_n(x)$ and $K_n(x)$ are Bessel functions of the first and second kind, respectively.

Near the Rindler horizon, i.e., for small $\rho\ll 1$, one has
\ba  \n{ASSIM}
&J_0(\omega \rho)\sim 1-{1\over 4}\omega^2 \rho^2\hh
J_0(\omega \rho)\sim {1\over 2}\omega \rho\, ,\\
&K_0(\omega \rho)\sim-\ln(\omega\rho)\hh K_1(\omega \rho)\sim{1\over \omega\rho}\, .
\ea

The electromagnetic field is regular at the Rindler horizon if its components are finite and smooth functions of $(T,X,Y,Z)$ coordinates. For the field (\ref{FRZ}), one has
\be \n{FTXZ}
F_{T Z}=-{T\over \rho} F_{\rho z}=-{CT\over \rho^2}\hhh F_{X Z}={CX\over \rho} F_{\rho z}={CX\over \rho^2}\, .
\ee
Hence for a regular at the Rindler horizon solution (\ref{FRZ}), one should put $C=0$ (such that $F_{\rho z}=0$).

For regularity of the field at the Rindler horizon, it is sufficient that the components of its vector potential  $A_T$, $A_X$, $A_Y$ and $A_Z$ are finite and smooth functions of $(T,X,Y,Z)$. Using (\ref{ASSIM}), one can see that for the solution $A_y$ given by (\ref{SOL}), one must put $C^2_{y}(\omega)=0$. Using (\ref{RRCC}), one can also find the component  $A_{\tau}$ of the vector potential in Cartesian $(T,X)$ coordinates
\be \n{carpot}
A_T={X\over \rho^2}A_{\tau}\hh A_X=-{T\over \rho^2}A_{\tau}\, ,
\ee
and for a regular field one should put $C^2_{\tau}(\omega)=0$ in  (\ref{SOL}).

In the near-horizon approximation, we choose the coordinate $z$ such that $z=0$ corresponds to the equatorial plane. One may use relations (\ref{SOL}) to find the expression for the regular electromagnetic field in the domain near the horizon and close to the equatorial plane, where $\rho\ll 1$ and $z\ll 1$.

However, for $A_y$ this can be done directly. Let us  denote
\be  \n{sumy}
P_k=P_k(\rho,z)\equiv \sum_{n=0}^k \sum_{i=0}^n C_{i,n-i} \rho^i z^{n-i}\, ,
\ee
and write the leading asymptotic of a regular near the horizon function $A_y$ in the form $A_y=P_2(\rho,z)$.
Substituting this expression in the Maxwell equation $J^y=0$, one finds
\be
C_{10}=C_{11}=0\hh C_{02}=-2C_{20}\, .
\ee
Thus a regular at the horizon field $A_y$ has the form
\be \n{ANSy}
A_y={1\over 2}b (\rho^2-2z^2)+b_1 z={1\over 2}b (X^2-T^2-2Z^2)+b_1 Z\, .
\ee
Here, we have set $C_{22}^y={1\over 2} b$ and $C_{10}=b_1$. We have also set $C_{00}=0$ since this constant does not enter the expression for the field strength. The final expression in (\ref{ANSy}) clearly shows that $A_y$ is regular at the Rindler horizon. The field $F_{\mu\nu}$ for this potential has the following components
\be \n{FFRy}
F_{\rho y}=b\rho\hh F_{zy}=-2bz+b_1\, .
\ee
The electric field in the Rindler frame for the potential $A_y$,
\be
E^{\mu}=F^{\mu\nu}u_{\nu}\equiv -\rho^2 F^{\mu\tau},
\ee
vanishes, while the magnetic field
\be
B^{\mu}={1\over 2} e^{\mu\nu\alpha\beta}u_{\nu}F_{\alpha\beta}\, ,
\ee
has the following non-vanishing components
\be
B^{\rho}=-2bz+b_1\hh B^z=b\rho\, .
\ee
For $b=0$, this magnetic field is orthogonal to the horizon. Such a field in the near-horizon region of the black hole exists if the black hole has a non-vanishing magnetic monopole charge. In the absence of the monopole charge, i.e., when $b_1=0$, the leading term of the near-horizon magnetic field has a universal form which contains only one constant, $b$. In the case of a black hole immersed in a homogeneous at infinity magnetic field, $b$ coincides with its value at infinity.

A similar analysis can be done for the $A_{\tau}$ component of the potential, which is responsible to the electric field.
We write it in the form
\be
A_{\tau} = \rho^2 {\cal A}(\rho, z)\, .
\ee
The electromagnetic field for this potential is regular at the horizon when ${\cal A}$ is regular function of the Minkowski coordinates.
The equation $J^{\tau}=0$ implies that this function satisfies the equation
\be\n{FFJT}
\partial_{\rho}^2{\cal A}(\rho, z) + {3\over\rho}\partial_{\rho}{\cal A}(\rho, z)+\partial_{z}^2{\cal A}(\rho, z)=0\, .
\ee
We again use an asymptotic form of ${\cal A}$, namely ${\cal A}=P_2(\rho,z)$, near the $z=0$ plane in the vicinity of the horizon. Then  equation (\ref{FFJT}) imposes the following conditions on the $C_{i,2-i}$
\be
C_{10} = C_{11} = 0 \hh C_{02}=-4C_{20}\, .
\ee
Thus
\be
A_{\tau} = \rho^2\left[C_{00} + C_{01}z + C_{20}\bigg( \rho^2-4z^2\bigg)\right],
\ee
which is clearly regular with respect to Cartesian coordinates.

The resulting nonzero components of the field tensor are
\ba\n{FEXP}
F_{\rho\tau} &= 2\rho\bigg[C_{00} + C_{01}z + C_{20}\bigg(2\rho^2-4z^2\bigg)\bigg],\\
F_{z\tau} &= \rho^2\bigg[C_{01} - 8C_{20}z\bigg]\, .
\ea

Using relations (\ref{RRCC}), one can show that
\be
{\pa \tau\over \pa T}{\pa \rho\over \pa X}-{\pa \tau\over \pa X}{\pa \rho\over \pa T}={1\over \rho}\, .
\ee
Thus one has
\be
F_{TX}={1\over \rho}F_{\tau\rho}\, ,
\ee
and the leading term in the expression for $F_{\rho\tau}$ in (\ref{FEXP}) with coefficient $C_{00}=E/2$ describes a constant electric field $\vec{E}$ in flat spacetime directed along the $X$-axis.

\hspace{2cm}

%


\end{document}